\documentclass[12pt]{iopart}
\usepackage{iopams}
\usepackage{amsthm}
\usepackage{nicefrac}
\newcommand\horthop{\widetilde{h}{}}

\newtheorem{thm}{Theorem}[section]

\newtheorem{prop}[thm]{Proposition}

\theoremstyle{definition}
\newtheorem{defn}[thm]{Definition}
\theoremstyle{remark}

\def \bh {\mbox{{\bf h}}}

\begin{document}


\title[Symmetric teleparallel geometries]{Symmetric teleparallel geometries}

\author {A. A. Coley$^1$, R. J. van den Hoogen$^2$ and D. D. McNutt$^3$}

\address{$^1$ Department of Mathematics and Statistics, Dalhousie University, Halifax, Nova Scotia, Canada, B3H 3J5}
\ead{aac@mathstat.dal.ca}

\address{$^2$ Department of Mathematics and Statistics, St. Francis Xavier University, Antigonish, Nova Scotia, Canada, B2G 2W5}
\ead{rvandenh@stfx.ca}

\address{$^3$ Department of Mathematics and Physics, University of Stavanger, Stavanger, Norway}
\ead{david.d.mcnutt@uis.no}



\begin{abstract}
In teleparallel gravity and, in particular, in $F(T)$ teleparallel gravity, there is a challenge in determining an appropriate (co-)frame and its corresponding spin connection to describe the geometry.  Very often, the ``proper'' frame, the frame in which all inertial effects are absent, is not the simplest (e.g, diagonal) (co-)frame. The determination of the frame and its corresponding spin connection for $F(T)$ teleparallel gravity theories when there exist affine symmetries is of much interest. In this paper we present the general form of the coframe and its corresponding spin connection for teleparallel geometries which are invariant under a $G_6$ group of affine symmetries. The proper coframe and the corresponding $F(T)$ field equations are also shown for these Teleparallel Robertson Walker (TRW) geometries.   Further, with the addition of an additional affine symmetry, it is possible to define a Teleparallel de Sitter (TdS) geometry.

\end{abstract}




\section{Introduction}

In teleparallel theories of gravity (summarized in the Appendix), the tetrad (or (co)frame) and corresponding spin-connection replace the metric as the principal object of study.
Often when people in the literature talk about symmetric spacetimes, they are actually talking  about metric-symmetric spacetimes, i.e, Killing vectors. However, in teleparallel geometries this may not be an appropriate approach and therefore motivates the further study of symmetries in teleparallel gravity.

An affine frame (intrinsic) symmetry on the frame bundle of $M$, is a diffeomorphism from the manifold to itself  which leaves the spin-connection invariant and affects the invariant frame in a very restricted manner, which is characterized by a vector field, ${\bf X}$, satisfying \cite{Coley:2019zld,McNutt_Coley_vdH2022}:
\begin{equation}
\mathcal{L}_{{\bf X}} \bh_a = \lambda_a^{~b} \,\bh_b \mbox{ and } \mathcal{L}_{{\bf X}} \omega^a_{~bc} = 0, \label{Intro:FS2}
\end{equation}
where $\omega^a_{~bc}$ denotes the spin-connection relative to the geometrically preferred invariant frame $\bh_a$ determined by the Cartan-Karlhede algorithm \cite{Coley:2019zld,Stephani,Fonseca-Neto:1992xln} and $\lambda_a^{~b}$ is an element of the linear isotropy group determined by the algorithm. How the matrix $\lambda_a^{~b}$ is chosen will be briefly discussed in the following section.  Details can be found in \cite{McNutt_Coley_vdH2022}. This definition is {\it a frame-dependent} analogue of the definition of a symmetry introduced by \cite{HJKP2018}.

The Cartan-Karlhede algorithm is a method originally developed to locally characterize a geometry uniquely in terms of a finite set of invariants \cite{Coley:2019zld,Stephani,Fonseca-Neto:1992xln}. In the context of this work, the algorithm uses the canonical forms of the torsion tensor and its covariant derivatives, up to some finite order, to determine a class of invariantly defined frames. Relative to this class of frames, the components of the torsion tensor and its covariant derivatives are invariants, called Cartan invariants. The Cartan-Karlhede algorithm also provides information on the dimension of the affine frame symmetry group, along with the dimension of its linear isotropy group.

Since any affine frame symmetry is an isometry but not all isometries are affine frame symmetries, it follows
that symmetries of a spacetime (affine symmetries) are more restrictive than “metric symmetries”.
Hence the set of isometries may not represent a set of intrinsic symmetries for a given teleparallel geometry.,
Spacetimes with a single affine frame symmetry, ${\bf X}$, were studied in \cite{ColeyvdH2022},   spacetimes with multiple affine frame symmetries having no isotropies in \cite{HMC} and spacetimes with isotropies in \cite{McNutt_Coley_vdH2022}.

For example, it is known  \cite{Coley:2019zld} that there are no teleparallel geometries admitting a maximal group of affine frame symmetries other than Minkowski space  \cite{HJKP2018}.  In particular, de Sitter and anti-de Sitter spaces are not maximally symmetric spaces in teleparallel geometry. In general, if a  $4$-dimensional Riemann-Cartan geometry admits a non-zero torsion tensor, then the maximum dimension of the group of affine symmetries is at most seven \cite{Coley:2019zld}. Indeed, in the example of de Sitter in teleparallel gravity (TdS) presented in \cite{Coley:2019zld}, neither of the two inequivalent frames displayed, along with the trivial connection and with the identical de Sitter metric, are maximally symmetric [i.e., affine invariant under the full $G_{10}$ group]. The first frame has an affine symmetry group of 7 dimensions
which acts on the 4D spacetime [i.e., the space is 4D homogeneous]. The second has a 4 dimensional symmetry group. In these two examples, one frame is diagonal while the other is not and so there is no Lorentz transformation mapping one frame into the other which also preserves the trivial connection condition. Hence the two teleparallel geometries are not related by a diffeomorphism. We note that it is the first teleparallel geometry that is a special case of the
teleparallel analogue of Robertson-Walker (TRW) geometry with a $G_6$ Lie algebra of affine symmetries (we will describe this below) .

Indeed, there is some confusion in discussions of TRW geometry in the literature. It is not always clear
whether the  frame and the spin connection considered has the full $G_6$ Lie algebra of affine symmetries consisting of spatially homogeneous and isotropic symmetries, which are isometries of the metric.  This is also relevant regarding the discussion of teleparallel de Sitter (TdS) geometries considered later. In particular, in the literature, when the parameter $k$ is non-zero,  the geometries presented do not always have a $G_6$ of affine frame symmetries; typically only 3 of the Killing vectors (KVs) are affine frame symmetries. When a proper frame is considered additional confusion is often present.

Let us comment on the parameter $k$ which determines the nature of the spatial curvature in the RW metric.  Since the Riemmann curvature is identically zero in teleparallel geometries, $k$ cannot be related to the spatial curvature of the Riemann tensor. While ``$\nicefrac{k}{a^2}$'' can still be interpreted as the curvature of a particular 3-space, in 4-space it is a component of the torsion scalar $T = 6(H^2-\nicefrac{k}{a^2})$.

The situation appears to be worse in the case $k = -1$, in which the absurd situation of complex valued components of the frame or spin connection are used \cite{Ferraro:2011us}. Indeed, for $k=-1$ the connection considered is not ideal for a real-valued teleparallel geometry. Frame/connection pairs for geometries satisfying a $G_6$ group of affine symmetries have been presented by Hohmann \cite{Hohmann:2015pva,Hohmann:2021ast} and others \cite{DAmbrosio:2021pnd,DAmbrosio:2021zpm}.  However, it is still not clear which functions in these presentations are essential to determining a solution and which are coordinate dependent. A second challenge is that these approaches cannot explicitly determine subclasses which admit additional symmetry, (instead larger symmetry groups must be assumed and determined to exist or not exist by trial and error). While most of the literature to date has concentrated on the analysis of cosmological models having a $k=0$ Robertson-Walker (RW) metric \cite{Bahamonde:2021gfp}, one can now begin expanding our understanding of how different symmetry assumptions affect the cosmological models built in teleparallel theories of gravity.

From the discussion above, we wish to revisit the ideas of affine symmetries. In particular, we seek to first determine all  teleparallel geometries which admit the full $G_6$ Lie algebra of affine symmetries. We will call such geometries teleparallel Robertson Walker (TRW) geometries.
Building on these results, we will then propose a definition for the teleparallel de Sitter (TdS) geometry.

\subsection{Notation}
The notation employed uses Greek indices $(\mu,\nu,\dots)$ to represent space-time coordinate indices and Latin indices $(a,b,\dots)$, to represent frame or tangent-space indices. Round brackets surrounding indices represents symmetrization, while square brackets represents anti-symmetrization. The frame basis is denoted as $h_a$ with the corresponding coframe basis $h^a$ where $h_a^{~\mu} h^b_{~\mu} = \delta^a_b$. The proper coframe is designated with a $\horthop^a$. The metric signature is $(-+++)$.


\section{ Teleparallel Robertson-Walker (TRW) spacetimes}

\subsection{Affine Symmetry Groups with non-trivial Isotropies}

Teleparallel geometries are characterised by a frame (or coframe) and a metric compatible, zero-curvature connection.  As a starting point to determine the TRW geometries, we will apply the following proposition \cite{McNutt_Coley_vdH2022}:

\begin{prop}
The most general teleparallel geometry which admits a given group of symmetries, ${\bf X}_I,~ I,J,K \in \{1,\dots,N\}$ with a non-trivial linear isotropy group of dimension $s$ can be determined by solving for the unknowns $h^a_{~\mu}$, $f_I^{~\hat{i}}$ (with $\hat{i}, \hat{j}, \hat{k} \in \{1,\dots,n\}$) and $\omega^a_{~bc}$ from the following equations:
\begin{eqnarray}
&& \mathcal{L}_{X_I} h^a_{~\mu} = f_I^{~\hat{i}} \lambda^a_{\hat{i}~b} h^b_{~\mu} \label{Sym:Frm}\\
&& 2X_{[I} ( f_{J]}^{~\hat{k}}) - f_I^{~\hat{i}} f_J^{~\hat{j}} C^{\hat{k}}_{~\hat{i} \hat{j}} = C^K_{~IJ} f_K^{~\hat{k}} \label{Sym:Com}\\
&& \mathcal{L}_{{\bf X}} \omega^a_{~bc} = 0, \label{Sym:SP} \\
&& h_c^{~\mu} \partial_\mu \omega^a_{\phantom{a}bd}- h_c^{~\nu}\partial_\nu \omega^a_{\phantom{a}bd}+\omega^a_{\phantom{a}fc}\omega^f_{\phantom{a}bd}-\omega^a_{\phantom{a}fd}\omega^f_{\phantom{a}bc} = 0.  \label{Sym:Riem}
\end{eqnarray}
where $\{ \lambda^a_{\hat{i}~b}\}_{\hat{i}=1}^s$ are a basis for the Lie algebra of the linear isotropy group, $[\lambda_{\hat{i}}, \lambda_{\hat{j}}] = C^{\hat{k}}_{~\hat{i}\hat{j}} \lambda_{\hat{k}}$, $[{\bf X}_I, {\bf X}_J] = C^K_{~IJ} {\bf X}_K$.
\end{prop}
\noindent Equations \eref{Sym:Frm} and \eref{Sym:SP} are the affine symmetry conditions, equation \eref{Sym:Com} results from the commutator property of Lie derivatives $[\mathcal{L}_{\mathbf{X}_I},\mathcal{L}_{\mathbf{X}_J}] = \mathcal{L}_{[\mathbf{X}_I,\mathbf{X}_J]}$ while equation \eref{Sym:Riem} is the zero curvature condition.

Working in coordinates $(t, r, \theta, \phi)$, the Killing vectors associated with the RW metric are:
\begin{eqnarray} & X_z = \partial_{\phi},~X_y = - \cos \phi \partial_{\theta} + \frac{\sin \phi}{\tan \theta} \partial_{\phi}, X_x = \sin \phi \partial_{\theta} + \frac{\cos \phi}{\tan \theta} \partial_{\phi}, \nonumber \\
& X_1 = \chi \sin \theta \cos \phi \partial_r + \frac{\chi}{r}\cos \theta \cos \phi \partial_{\theta} - \frac{\chi \sin \phi}{r \sin \theta} \partial_{\phi}, \nonumber \\
& X_2 = \chi \sin \theta \sin \phi \partial_r + \frac{\chi}{r} \cos \theta \sin \phi \partial_{\theta} + \frac{\chi \cos \phi}{r \sin \theta} \partial_{\phi}, \label{G6 generators} \\
& X_3 = \chi \cos \theta \partial_r - \frac{\chi}{r} \sin \theta \partial_{\theta}. \nonumber
\end{eqnarray}
where $\chi=\sqrt{1-kr^2}$.  We write $\{ X_I\}_{I=1}^6 = \{ X_1, X_2, X_3, X_x, X_y, X_z\}$, with the commutator constants, $C^I_{~JK}$.

The largest linear isotropy group permitted by a spatially homogeneous geometry is $SO(3)$ and a matrix basis for its Lie algebra is of the form:
\begin{equation}\fl
 \lambda_{\hat{1}} = \left[ \begin{array}{cccc} 0 & 0 & 0 & 0 \\
0 & 0 & 0 & 0 \\ 0 & 0 & 0 & 1 \\ 0 & 0 & -1 & 0 \end{array} \right],~ \lambda_{\hat{2}} = -\left[ \begin{array}{cccc} 0 & 0 & 0 & 0 \\
0 & 0 & 1 & 0 \\ 0 & -1 & 0 & 0 \\ 0 & 0 & 0 & 0 \end{array} \right] ,~
\lambda_{\hat{3}} =  -\left[ \begin{array}{cccc} 0 & 0 & 0 & 0 \\
0 & 0 & 0 & 1 \\ 0 & 0 & 0 & 0 \\ 0 & -1 & 0 & 0 \end{array} \right]  \label{so3LA}
\end{equation}
with the corresponding commutator constants, $C^{\hat{i}}_{~\hat{j} \hat{k}}$: $C^{\hat{3}}_{~\hat{1} \hat{2}} = -C^{\hat{2}}_{~\hat{1} \hat{3}} = ~C^{\hat{1}}_{~\hat{2} \hat{3}} = -1 $.

The first step is to solve equation \eref{Sym:Com} and determine the form of the functions in $f_I^{~\hat{i}}$ through:
\begin{equation}
2X_{[I} ( f_{J]}^{~\hat{k}}) - f_I^{~\hat{i}} f_J^{~\hat{j}} C^{\hat{k}}_{~\hat{i} \hat{j}} = C^K_{~IJ} f_K^{~\hat{k}}. \label{FLRWstart}
\end{equation}
We exploit the freedom of choice in the components of $f_I^{~\hat{i}}$ using the isotropy group. The isotropy group affects a change to these components through the equation:
\begin{equation}
X_I(\tilde{\Lambda}^a_{~b}) [\tilde{\Lambda}^{-1}]^b_{~c} + \tilde{\Lambda}^a_{~b} f_I^{~\hat{i}} \lambda^b_{\hat{i}~e} [\tilde{\Lambda}^{-1}]^e_{~c} = {f}_I^{~\hat{j}} \lambda_{\hat{j}~a}^{~b} .\label{ChangeIso}
\end{equation}
where $\tilde{\Lambda}^a_{~b}$ is some element of the isotropy group.
Since $X_z$ is a generator of a spatial rotation, we will choose our frame representation so that it acts as a rotation on the basis elements $\bh^3$ and $\bh^4$, i.e.,
$f_6^{~\hat{1}}=f_6^{~\hat{2}}=0$.
By applying a rotation about $\bh^3$ and $\bh^4$, this remaining component $f_6^{~\hat{3}}$ can be set to zero using
\begin{equation}
 X_6(\tilde{\Lambda}^a_{~b}) [\tilde{\Lambda}^{-1}]^b_{~c} + \tilde{\Lambda}^a_{~b} f_6^{~\hat{i}} \lambda^b_{\hat{i}~e} [\tilde{\Lambda}^{-1}]^e_{~c} = 0
\end{equation}
to set $f_6^{~\hat{3}} = 0$. Equation \eref{Sym:Com} can then be solved in a straightforward manner by using the remaining freedom in the isotropy group and equation \eref{ChangeIso}:
\begin{equation}
f_I^{~\hat{i}} = \left [ \arraycolsep=3pt\def\arraystretch{1.5}
\begin{array}{ccc} - \frac{\sqrt{1-kr^2} \sin(\phi) \cos(\theta)}{r \sin(\theta)} & -\frac{\cos(\theta) \cos(\phi)}{r} & \frac{\sin(\phi)}{r} \\ \frac{\sqrt{1-kr^2} \cos(\phi) \cos(\theta)}{r \sin(\theta)} & -\frac{\cos(\theta) \sin(\phi)}{r} & - \frac{\cos(\phi)}{r} \\ 0 & \frac{\sin(\theta)}{r} & 0 \\ \frac{\cos(\phi)}{\sin(\theta)} & 0 & 0 \\ \frac{\sin(\phi)}{\sin(\theta)} & 0 & 0 \\ 0 & 0 & 0 \end{array} \right]
\end{equation}
With the representation of the linear isotropy group's Lie algebra given in \eref{so3LA} and the form of $f_I^{~\hat{i}}$ determined we can solve equation \eref{Sym:Frm},
\begin{equation}
X_I^{~\nu} \partial_{\nu} h^a_{~\mu} + \partial_{\mu} X_I^{~\nu} h^a_{~\nu} = f_I^{~\hat{i}} \lambda^a_{\hat{i}~b} h^b_{~\mu}
\end{equation}
to determine the most general frame admitting this symmetry group.
In general, a coordinate transformation can be made to simplify the coframe.  This
coordinate transformation leads to a Lorentz transformation that will be absorbed by the connection.
Therefore, we can work with the following coframe:
\begin{equation}
h^a_{~\mu} = \left[ \begin{array}{cccc} 1 & 0 & 0 & 0 \\ 0 & \frac{a(t)}{\sqrt{1-kr^2}} & 0 & 0 \\ 0 & 0 & a(t) r & 0 \\ 0 & 0 & 0 & a(t) r \sin(\theta) \end{array} \right]. \label{VB:FLRW}
\end{equation}

Using the coframe associated to this matrix, we can readily solve the resulting equations coming from equation \eref{Sym:SP} for each of the Killing vector fields. We will assume that the connection is metric compatible, so that $\omega_{abc} = -\omega_{bac}$. The solution to these equations contains two arbitrary functions $W_1(t)$ and $W_2(t)$ and has the following non-trivial components:
\begin{eqnarray}
 & \omega_{122} = \omega_{133} = \omega_{144} =  W_1(t), \nonumber \\
& \omega_{234} = -\omega_{243} = \omega_{342} = W_2(t), \nonumber \\
& \omega_{233} = \omega_{244} = - \frac{\sqrt{1-kr^2}}{a(t)r}, \label{Con:FLRW}\\
& \omega_{344} =  - \frac{\cos(\theta)}{a(t) r \sin(\theta)}.  \nonumber
\end{eqnarray}
The above connection is the most general connection for any Riemann-Cartan geometry which admits the symmetry group with generators given by equation  \eref{G6 generators}. For this class of connections, the tensor-part of the torsion tensor is automatically zero and the torsion tensor can be decomposed into the vector-part, $V_a = T^b_{~ba}$ and axial-part $A_a = \frac{1}{6}\epsilon_{abcd} T^{bcd}$.

To determine the class of connections that describe a teleparallel geometry we must impose the flatness condition in equation \eref{Sym:Riem}.
This has a number of distinct solutions and leads to the following proposition

\begin{prop}

The class of TRW geometries are given by the frame \eref{VB:FLRW} and connection \eref{Con:FLRW} are split into a number of distinct cases, determined by the arbitrary functions $W_1$ and $W_2$ in the connection.
\begin{itemize}
\item $W_1(t) = 0$, $\displaystyle W_2(t) = \pm \frac{\sqrt{k}}{a(t)}$ where $\displaystyle {\bf V} = - \frac{3\dot{a}(t)}{a(t)} \bh^1, \mbox{\ \rm and\ } {\bf A} = \mp \frac{2\sqrt{k}}{a(t)} \bh^1$

\item $\displaystyle W_1(t) = \pm \frac{\sqrt{-k}}{a(t)}$, $W_2(t) = 0$ where $\displaystyle{\bf V} = \pm \frac{3(\sqrt{-k} + \dot{a}(t))}{a(t)} \bh^1,\mbox{\ \rm and\ } {\bf A} = 0 $
\end{itemize}
\noindent where $a(t)$ is the frame function and $\dot{a}(t)$ is its derivative. Each case above contains the subcase $k=0$.
\end{prop}

In conclusion, the coframe \eref{VB:FLRW} with the connection \eref{Con:FLRW} will be a teleparallel geometry admitting the desired symmetry group with generators given by equation \eref{G6 generators}, if the two arbitrary functions in the connection satisfy one of the above forms. It is natural to ask that these functions are real-valued, and this constraint immediately distinguishes the RW metrics with positive and negative $k$.

\subsection{TRW Proper Frames}

Assuming that the spacetime geometry is invariant under a $G_6$ group of affine frame symmetries, McNutt et al. \cite{McNutt_Coley_vdH2022} determined the form of the corresponding spin connection briefly described above.  Assuming an orthonormal diagonal coframe in spherical coordinates \eref{VB:FLRW} with corresponding spin connection determined by equation \eref{Con:FLRW} we can consider the three scenarios for $k$.
For $k=-1$ we have $W_1(t)=\delta\sqrt{-k}/a(t)$ and $W_2(t) =0$, for $k=0$ we have $W_1(t)=0$ and $W_2(t)=0$, and for $k=+1$ we have that $W_1(t)=0$ and $W_2(t)=\delta\sqrt{k}/a(t)$  where in each case there is a discrete parameter $\delta=\pm1$.
Since there exists a matrix $\Lambda^a_{~b}\in SO(1,3)$ that yields the spin connection via the differential equation
\begin{equation}
\omega^a_{~b}=(\Lambda^{-1})^a_{~c}d\Lambda^c_{~b}\label{DE_SPIN}
\end{equation}
all that needs to be done is to solve this system of differential equations for $\Lambda^a_{~b}$ in each of the situations $k=-1$ and $k=+1$ (noting that $k=0$ is a subcase of both).  With this $\Lambda^a_{~b}$ one can easily determine a proper coframe $\horthop^a = \Lambda^a_{~b}h^b$, where $h^b$ is given by equation \eref{VB:FLRW}.

The torsion scalar for any of the scenarios $k=(-1,0,1)$ has the form
\begin{eqnarray}
T(t) &=& 6\left(\frac{\dot{a}}{a}\right)^2 + 12W_1\frac{\dot{a}}{a}+6W_1^{\,2}-6W_2^{\,2}\\
      &=& 6\left(\frac{\dot{a}}{a}+W_1+W_2\right)\left(\frac{\dot{a}}{a}+W_1-W_2\right)\\
      &=& -\frac{2}{3}V^2 + \frac{3}{2}A^2
\end{eqnarray}
where, the magnitudes of the vectorial and axial terms are
\begin{equation*}
V^2=-9\left(\frac{\dot{a}}{a}+W_1\right)^2, \qquad A^2 = -4W_2^2.
\end{equation*}
It is curious to note how a non-trivial axial part of the torsion scalar only appears in the $k=+1$ case.

\subsubsection{Negative $k$-parameter TRW Case:}

When $k=-1$, a Lorentz transformation that satisfies the differential equation \eref{DE_SPIN} is
\begin{equation}\fl
\Lambda^a_{~b} = \left[\arraycolsep=3pt\def\arraystretch{1.5} \begin{array}{cccc}
    \sqrt{1-kr^2}&                  -\delta\sqrt{-k} r&                                  0&                             0\\
    -\delta\sqrt{-k} r\sin(\theta)\sin(\phi)&   \sqrt{1-kr^2}\sin(\theta)\sin(\phi)&    \cos(\theta)\sin(\phi)&          \cos(\phi)\\
     -\delta\sqrt{-k} r \sin(\theta)\cos(\phi)&   \sqrt{1-kr^2}\sin(\theta)\cos(\phi)&    \cos(\theta)\cos(\phi)&         -\sin(\phi)\\
      \delta\sqrt{-k} r \cos(\theta)&           -\sqrt{1-kr^2}\cos(\theta)&             \sin(\theta)&                   0
\end{array}\right].
\end{equation}
We note that any global Lorentz transformation multiplying this transformation is also a solution.
Therefore, one option to properly formulate the $k=-1$ spacetime geometry is to use the proper frame $\horthop^a = \Lambda^a_{~b}h^b$ with this Lorentz transformation and with $h^b$ given by equation \eref{VB:FLRW},
which necessarily has a trivial spin connection.
A second choice is to use the diagonal coframe \eref{VB:FLRW} and corresponding spin connection one-form
\begin{equation}\fl
\omega^a_{~b} =\left[\arraycolsep=3pt\def\arraystretch{1.5} \begin{array}{cccc}
 0                                & -\frac{\delta\sqrt{-k}}{\sqrt{1-kr^2}}dr   & -\delta\sqrt{-k} r d\theta          & -\delta\sqrt{-k} r\sin(\theta) d\phi \\
 -\frac{\delta\sqrt{-k}}{\sqrt{1-kr^2}}dr   & 0                                & -\sqrt{1-kr^2}d\theta       & -\sqrt{1-kr^2}\sin(\theta)d\phi \\
 -\delta\sqrt{-k} r d\theta                & \sqrt{1-kr^2}d\theta              & 0                          & -\cos(\theta)d\phi \\
 -\delta\sqrt{-k} r\sin(\theta) d\phi      & \sqrt{1-kr^2}\sin(\theta)d\phi    & \cos(\theta)d\phi         & 0
\end{array}\right].\label{conn_neg}
\end{equation}

Using either the proper coframe (and trivial connection) or the diagonal coframe/connection pair \eref{VB:FLRW} and \eref{conn_neg} the torsion scalar is
\begin{equation}
T=6\left( \frac{\dot{a}}{a}+ \frac{\delta\sqrt{-k}}{a}\right)^2.
\end{equation}
We assume an energy momentum tensor of the form $T_{ab}=\rho(t)u_a u_b +(u_a u_b+g_{ab})p(t)$ describing a perfect fluid with energy density $\rho(t)$ and pressure $p(t)$.
The antisymmetric part of the $F(T)$ teleparallel gravity field equations \eref{temp2} are identically satisfied and the linearly independent field equations from the symmetric part of the $F(T)$ teleparallel gravity field equations \eref{temp1} are
\begin{eqnarray}
\fl -\frac{F(T)}{2}+6F'(T)\left(\frac{\dot{a}}{a}\right)\left(\frac{\dot{a}}{a}+\frac{\delta\sqrt{-k}}{a}\right)=\kappa\rho,\\
\fl F(T)-6F'(T)\left(\frac{\ddot{a}}{a}+\left(\frac{\dot{a}}{a}+\frac{\delta\sqrt{-k}}{a}\right)^2 \right)-6F''(T)\dot{T}\left(\frac{\dot{a}}{a}+\frac{\delta\sqrt{-k}}{a}\right)=\kappa(\rho+3p).
\end{eqnarray}
The energy conservation equation that follows from the above equations is
\begin{equation}
\dot\rho = -3\frac{\dot{a}}{a}(\rho + p). \label{Energy_Conservation}
\end{equation}
These coframe connection pairs are also valid in the subcase $k=0$. Notice that we cannot use this construction when $k=+1$ because complex valued coframes or spin connections result.

\subsubsection{Positive $k$-parameter TRW Case:}

When $k=+1$, a Lorentz transformation that satisfies the differential equation \eref{DE_SPIN} is
\small\begin{equation}\fl
\Lambda^a_{~b} = \left[\arraycolsep=3pt\def\arraystretch{1.5} \begin{array}{cccc}
    1&                  0       &                                  0                 &                 0\\
    0&   \cos(\theta)           &    -\sqrt{1-kr^2}\sin(\theta)                       &     -\delta\sqrt{k}r\sin(\theta)\\
    0&   \sin(\theta)\cos(\phi) &     \sqrt{1-kr^2}\cos(\theta)\cos(\phi)-\delta\sqrt{k}r\sin(\phi) &     \delta\sqrt{k}r\cos(\theta)\cos(\phi)-\sqrt{1-kr^2}\sin(\phi)\\
    0&   \sin(\theta)\sin(\phi) &     \sqrt{1-kr^2}\cos(\theta)\sin(\phi)+\delta\sqrt{k}r\cos(\phi) &     \delta\sqrt{k}r\cos(\theta)\sin(\phi)+\sqrt{1-kr^2}\cos(\phi)
\end{array}\right].
\end{equation}\normalsize
Again any global Lorentz transformation multiplying this transformation is also a solution.

An option to properly formulate the $k=+1$ geometry is to construct the proper frame $\horthop^a = \Lambda^a_{~b}h^b$ using this Lorentz transformation and with $h^b$ given by equation \eref{VB:FLRW},
which necessarily has a trivial spin connection. A second choice is to use the diagonal coframe \eref{VB:FLRW} and corresponding spin connection one-form
\small\begin{equation}\fl
\omega^a_{~b} =\left[\arraycolsep=3pt\def\arraystretch{1.5} \begin{array}{cccc}
 0   & 0                                        & 0                                          & 0 \\
 0   & 0                                        &  -\sqrt{1-kr^2}d\theta +\delta\sqrt{k}r\sin(\theta)d\phi        & -\delta\sqrt{k}rd\theta -\sqrt{1-kr^2}\sin(\theta)d\phi \\
 0   & \sqrt{1-kr^2}d\theta -\delta\sqrt{k}r\sin(\theta)d\phi        & 0                                          & \frac{\delta\sqrt{k}}{\sqrt{1-kr^2}}dr-\cos(\theta)d\phi \\
 0   & \delta\sqrt{k}rd\theta +\sqrt{1-kr^2}\sin(\theta)d\phi & -\frac{\delta\sqrt{k}}{\sqrt{1-kr^2}}dr+\cos(\theta)d\phi & 0
\end{array}\right].
\end{equation} \normalsize

Using either the proper coframe (and trivial spin connection) or the diagonal coframe/connection pair \eref{VB:FLRW} and \eref{conn_neg} the torsion scalar is
\begin{equation}
T=6\left( \left(\frac{\dot{a}}{a}\right)^2- \frac{k}{a^2}\right)
\end{equation}
which is independent of $\delta$.
Assuming a perfect fluid as in the $k=-1$ case, the antisymmetric part of the field equations are identically satisfied and the symmetric part of the field equations are
\begin{eqnarray}
-\frac{F(T)}{2}+6F'(T)\left(\frac{\dot{a}}{a}\right)^2=\kappa\rho,\\
F(T)-6F'(T)\left(\frac{\ddot{a}}{a}+\left(\frac{\dot{a}}{a}\right)^2-\frac{k}{a^2}
\right)-6F''(T)\dot{T}\left(\frac{\dot{a}}{a}\right)=\kappa(\rho+3p).
\end{eqnarray}
In this $k=+1$ case, there is no dependence on the discrete parameter  $\delta$.
The energy conservation equation is again given by \eref{Energy_Conservation}.
Further, the equations once again reduce to the $k=0$ field equations by setting $k=0$ or become invalid if $k=-1$.

\subsubsection{Discussion}

In most cases involving the construction of cosmological models in $F(T)$ teleparallel gravity, authors have only considered the  $k=0$ case (e.g., see review  \cite{Cai_2015}). For the $k=\pm 1$ cases, there have been many erroneous attempts to find solutions, many involving the use of complex tetrads.  Ferraro and Fiorini \cite{Ferraro:2011us}  made one of the first attempts to properly determine the frame corresponding to a particular assumed symmetry. They were successful in constructing a proper frame for the $k=+1$ case, but the corresponding $k=-1$ case resulted in a complex quantities. More recently, Hohmann and collaborators, were successful in constructing real valued proper frames for both the $k=-1$ and $k=+1$ cases \cite{HJKP2018,Hohmann:2018rwf,Pfeifer:2022txm}.  The geometrical approaches employed in \cite{HJKP2018,Hohmann:2018rwf,Pfeifer:2022txm} are similar to that presented here.  However, in short, the definitions are not identical, we only discuss affine symmetries, and we avoid any complex quantities. Since we have made no assumptions a priori, our results are general. We shall discuss this in more detail in \cite{McNutt_Coley_vdH2022} (which contains all the necessary mathematical details and some additional examples).


\section{Teleparallel ``\lowercase{de} Sitter'' (T\lowercase{d}S)}

There are no teleparallel geometries admitting a maximal group of affine frame symmetries other than Minkowski space  \cite{HJKP2018}.  If a  $4$-dimensional teleparallel geometry has a non-zero torsion, then the maximum dimension of the group of affine symmetries is at most seven \cite{Coley:2019zld}.  Therefore, let us investigate the following particular scenario to study the analogue of de Sitter geometries in GR.

Using the Cartan-Karlhede algorithm \cite{Coley:2019zld}, we can determine two different classes of $G_7$ geometries by requiring that the Cartan invariants are all constant. This follows from the formula for the dimension of the affine frame symmetry group,
$N = s+4-t_p$ where $s$ is the dimension of the linear isotropy group and $t_p$ is the number of functionally independent invariants at the conclusion of the algorithm. If all of the Cartan invariants are constant, then $t_p = 0$ and the dimension of the linear isotropy group is three, yielding a seven-dimensional affine frame symmetry group.

Solving the differential equations arising from the requirement that the only non-trivial components of the Cartan invariants are $T_{abc}$ gives two classes of possibilities. In each case we will use equations \eref{Sym:Frm}-\eref{Sym:SP} to determine the form of the new affine frame symmetry.

In the first case we have that $a(t) = A_0 e^{H_0 t}$, $k=0$, $H_0\not =0$ where $H_0$ is a constant. In this case, the affine frame symmetry is of the form
\begin{equation}
X_7 = -\frac{1}{A_0 H_0} \partial_t + r \partial_r.  \label{X7_desitter}
\end{equation}
The resulting Lie algebra of $\{X_I\}_{I=1}^{7} = \{X_1,X_2,X_3, X_x, X_y, X_z, X_7\}$ is given by
\begin{equation}
\begin{array}{lll}
{}  [X_1,X_5]=X_3,  & [X_1,X_6]=X_2,  & [X_1,X_7]=X_1, \\
{} [X_2,X_4]=-X_3, & [X_2,X_6]=-X_1, & [X_2,X_7]=X_2, \\
{}  [X_3,X_4]=X_2,  & [X_3,X_5]=-X_1, & [X_3,X_7]=X_3, \\
{}  [X_4,X_5]=-X_6, & [X_4,X_6]=X_5,  & [X_5,X_6]=-X_4.
\end{array} \nonumber
\end{equation}
By inspection, this is a subalgebra of the Lie algebra for the group of metric (Killing) symmetries of de Sitter spacetime.
We therefore propose the following definition.
\begin{defn}
The Teleparallel de Sitter geometry (TdS) is a teleparallel geometry with a $G_7$ Lie group of affine symmetries which is a subgroup of $O(1,4)$.
\end{defn}
\noindent Note in this geometry the covariant derivative of the torsion tensor is zero.

Using our coframe/spin connection pair determined earlier (either proper or not), the torsion scalar, and magnitudes of the vectorial and axial parts of the Torsion scalar for the TdS geometries are
\begin{equation}\arraycolsep=5pt\def\arraystretch{1.5} \begin{array}{cccc}
k=0,  & T=6H_0^{\,2}, & V^2=-9H_0^{\,2}, & A^2=0. \\
\end{array}
\end{equation}

The field equations in the TdS case reduce to
\begin{equation}
\kappa\rho = -\kappa p = -\frac{1}{2}F(T_0) + 6F'(T_0)H_0^{\,2}
\end{equation}
where necessarily $\rho$ and $p$ are constant.  The equations formally reduce to their GR counterparts only when $F(T)=T$. We note that the effective equation of state
\begin{equation}
\omega_{eff}=\frac{p}{\rho}=-1
\end{equation}
is the same as its GR counterpart, however, the effective cosmological constant $\Lambda_{eff}\equiv\kappa \rho$ depends on  the two parameters $F(T_0)$ and $F'(T_0)$.


\subsection{Teleparallel Einstein Static (TES)}

In the second case, $a(t) = A_0$ a non-zero constant, $k=\pm1$.  These geometries correspond to the direct product $\mathbb{R} \times M_3$, where $M_3$ is a locally homogeneous and isotropic Riemannian manifold. This is reflected in the Lie algebra structure of the affine frame symmetries  $\{X_I\}_{I=1}^{7} = \{X_1,X_2,X_3, X_x, X_y, X_z, X_7\}$, where $X_7 = \partial_t$ and
\begin{equation}
 \begin{array}{lll}
{} [X_1,X_5]=X_3, & [X_1,X_6]=X_2,& [X_2,X_4]=-X_3, \\
{} [X_2,X_6]=-X_1,& [X_3,X_4]=X_2,& [X_3,X_5]=-X_1, \\
{} [X_4,X_5]=-X_6,& [X_4,X_6]=X_5,& [X_5,X_6]=-X_4, \\
\multicolumn{3}{l}{[X_i, X_7] = 0, i \in \{1,\dots,6\}.}
\end{array} \nonumber
\end{equation}
Since $X_7 = \partial_t$ is the additional affine frame symmetry, this geometry is necessarily static.  This geometry can be considered as the analogue of the Einstein static geometry in GR, which we shall call the Teleparallel Einstein Static (TES) geometry.

Using our coframe/spin connection pair determined earlier (either proper or not), the torsion scalar, and magnitudes of the vectorial and axial parts of the torsion scalar are
\begin{equation}
\arraycolsep=5pt\def\arraystretch{1.5} \begin{array}{llll}
k=-1,   & T=-\frac{6k}{A_0^{\,2}},  & V^2=\frac{9k}{A_0^{\,2}}, &  A^2=0, \\
k=+1,   & T=-\frac{6k}{A_0^{\,2}},  & V^2=0, & A^2=-\frac{4k}{A_0^{\,2}}.
\end{array}
\end{equation}

The field equations in the TES case reduce to
\begin{equation}
\kappa\rho = -\frac{1}{2}F(T_0) \mbox{\ \ \rm and\ \ } \kappa p = \frac{1}{2}F(T_0)  + 2F'(T_0)\frac{k}{A_0^{\,2}}
\end{equation}
where necessarily $\rho$ and $p$ are constant.  For the TES model the effective equation of state
\begin{equation}
\omega_{eff}=\frac{p}{\rho}=-1-\frac{F'(T_0)}{F(T_0)}\frac{4k}{A_0^{\,2}},
\end{equation}
where in GR the Einstein static metric yields $\omega_{eff}=-1/3$.


\section{Discussion}

We have clarified the role of choosing an appropriate spin connection for a given coframe ansatz.  Using an algorithm developed in \cite{McNutt_Coley_vdH2022} we have constructed a set of invariant coframes and more importantly their corresponding spin connection that respects the affine frame symmetries that have been imposed.  In particular, we have presented the coframe and its corresponding spin connection for teleparallel geometries which are invariant under a $G_6$ group of affine symmetries.  In addition, the proper coframe has also been determined in each case and the field equations expressed.  The corresponding metric is Robertson-Walker type and is characterized by a spatial curvature parameter $k = \{-1,0,1\}$.  It is interesting to note that in the $k=-1$ case the field equations result in two different situations.

In the TRW cases, having an appropriate spin connection/coframe pair results in a situation in which the antisymmetric part of the field equations are identically zero.  Having an appropriate spin connection/coframe pair defined via symmetry requirements, is not always compatible with the antisymmetric part of the $F(T)$ teleparallel field equations.  Indeed it has been shown in \cite{ColeyvdH2022} that when there is a single affine symmetry requirement, that the antisymmetric part of the field equations place severe constraints on the geometry.

In  teleparallel geometries with non trivial torsion, since the dimension of the maximal group of affine symmetries is seven, we define the teleparallel de Sitter (TdS) geometry as that nontrivial teleparallel geometry which has a seven dimensional group of symmetries that is also a subgroup of the group of the Killing symmetries of the de Sitter metric. Furthermore, using a similar technique, we are able to define the teleparallel analogue of Einstein static geometry.

With the proposed definition of TdS, it now becomes possible to extend this work by considering perturbations of TdS.  Further, this analysis provides a solid foundation for the development of generalizations of TRW cosmological models through the construction and analysis of teleparallel Bianchi geometries.


\ack
AAC and RvdH are supported by the Natural Sciences and Engineering Research Council of Canada. RvdH is supported by the St Francis Xavier University Council on Research.

\appendix
\section{Appendix: Overview of $F(T)$ Teleparallel Gravity}

As an alternative to Riemmannian geometries which are typically characterized by the curvature of a Levi-Civita connection calculated from the metric, Teleparallel geometries are characterized by the torsion.  The torsion is a function of the coframe, derivatives of the coframe, and a zero curvature and metric compatible spin connection. Teleparallel geometries provide an alternative framework in which to build a theory of gravity.  A variety of teleparallel gravitational theories based on a Lagrangian can be constructed using various scalars built from the torsion and functions thereof.
One subclass of teleparallel gravitational theories is dynamically equivalent to GR and is appropriately called the Teleparallel Equivalent to General Relativity (TEGR) \cite{Aldrovandi_Pereira2013}

A particularly interesting generalization of the TEGR is $F(T)$ teleparallel gravity \cite{Ferraro:2006jd, Ferraro:2008ey, Linder:2010py}. In the {\it covariant} approach to $F(T)$ teleparallel gravity  \cite{Krssak:2018ywd}, the teleparallel geometry is defined in a gauge invariant manner as a geometry with zero curvature, having a spin-connection that vanishes in a very special class of frames (``proper frames'') where all inertial effects are absent, and non-zero in all other frames \cite{Aldrovandi_Pereira2013,Krssak:2018ywd,Lucas_Obukhov_Pereira2009}. Therefore, the resulting teleparallel gravity theory has Lorentz covariant field equations and is therefore locally Lorentz invariant \cite{Krssak_Pereira2015}.

As we have complete freedom to choose a coframe in which to do our computations, we choose the coframe so that the tangent space metric has the following form $g_{ab}= \eta_{ab} = \mbox{Diag}[-1,1,1,1]$.  This {\it orthonormal gauge} choice still allows a $O(1,3)$ subgroup of $GL(4,\mathbb{R})$ of residual gauge transformations which leaves the metric $g_{ab}=\eta_{ab}$ invariant. One can now restrict attention to the proper or proper ortho-chronous Lorentz subgroups, $SO(1,3)$ or $SO(1,3)^+$, as desired in any given physical situation.  Most importantly, within this orthonormal gauge choice, the resulting field equations transform homogeneously under the remaining $O(1,3)$ (or $SO(1,3)$ or $SO(1,3)^+$)  Lorentz gauge transformations.

The Lagrangian for $F(T)$ teleparallel gravity is given in terms of the scalar quantity, $T$, called the torsion scalar, defined in terms of the torsion tensor and the super-potential
\begin{equation}
T=\frac{1}{2}T^a_{\phantom{a}\mu\nu}S_a^{\phantom{a}\mu\nu}.
\end{equation}
where the super-potential, $S^a_{~\mu \nu}$, is constructed from the torsion tensor and various contractions with the metric and coframes,
\begin{equation}
S_a^{\phantom{a}\mu\nu}=\frac{1}{2}\left(T_a^{\phantom{a}\mu\nu}+T^{\nu\mu}_{\phantom{\nu\mu}a}
    -T^{\mu\nu}_{\phantom{\mu\nu}a}\right)-h_a^{\phantom{a}\nu}T^{\phi\mu}_{~~\phi} + h_a^{\phantom{a}\mu}T^{\phi\nu}_{~~\phi}. \label{super}
\end{equation}
The complete Lagrangian for $F(T)$ teleparallel gravity is
\begin{equation}
L= \frac{h}{2\kappa}F(T)+L_{Matt} \label{lagrangian}
\end{equation}
where $\kappa$ is the gravitational coupling constant, $\kappa=8\pi G/c^4$, where we have chosen units so that $c=1$.

The variations of the  Lagrangian, which include a non-trivial spin-connection \cite{Krssak_Saridakis2015,Krssak:2018ywd}, yield {\it Lorentz covariant} field equations.
If we consider the spin-connection as an independent quantity having zero curvature and being metric compatible, then the gravitational Lagrangian \eref{lagrangian} can be written using Lagrange multipliers to impose these two constraints. The corresponding variations yield the following
\begin{equation}
\omega^a_{\phantom{a}b\mu} = \Lambda^a_{\phantom{a}c}\partial_\mu\Lambda_{b}^{\phantom{a}{c}} \ \mbox{and}\ \omega_{(ab)\mu} = 0 \label{solution_omega}
\end{equation}
where given our orthonormal gauge choice, $\Lambda^a_{\phantom{a}b}\in SO(1,3)$  (Note: $\Lambda_b^{\phantom{b}c}
\equiv (\Lambda^{-1})^c_{\phantom{c}b}$).

We define the canonical energy momentum
\begin{equation}
h\Theta_a^{\phantom{a}\mu}=-\frac{\delta L_{Matt}}{\delta h^{\vphantom{\mu}a}_{\phantom{a}\mu}}.
\end{equation}
Since we have assumed the invariance of the field equations under $SO(1,3)$, the canonical energy momentum is symmetric $\Theta_{[ab]}=0$.
Further, it can be shown that the metrical energy momentum $T_{ab}$ now satisfies
\begin{equation}
T_{ab}\equiv-\frac{1}{2}\frac{\delta L_{Matt}}{\delta g_{ab}}=\Theta_{(ab)}.
\end{equation}

Variations of the Lagrangian describing $F(T)$ gravity \eref{lagrangian} with respect to the coframe can be decomposed into a  symmetric and antisymmetric parts
\begin{eqnarray}
\kappa \Theta_{(ab)}=
        F''(T)S_{(ab)}^{\phantom{(ab)}\nu} \partial_v T+F'(T)G_{ab}  + \frac{1}{2}g_{ab}\left(F(T)-TF'(T)\right),\label{temp1}\\
             0      = F''(T)S_{[ab]}^{\phantom{[ab]}\nu} \partial_v T,\label{temp2}
\end{eqnarray}
where $G_{ab}$ is the usual Einstein tensor calculated from the metric.

If $T = const.$, then the field equations for $F(T)$ teleparallel gravity are equivalent to a rescaled version of TEGR (which looks like GR with a cosmological constant and a rescaled coupling constant) \cite{Krssak:2018ywd}. In the case of TEGR, where $F(T)=T$, equation \eref{temp2} vanishes.  For $F(T) \neq T$, the variation of the gravitational Lagrangian by the flat spin connection is equivalent to the anti-symmetric part of the field equations in equation \eref{temp2} \cite{Krssak:2018ywd,Golovnev:2017dox}.  Further, since the canonical energy momentum is symmetric, $\Theta_{[ab]} = 0,$ then the anti-symmetric part of the field equations \eref{temp2} limit the possible solutions of spacetimes with a specific symmetry \cite{ColeyvdH2022}.


\section*{References}

\bibliographystyle{iopart-num.bst}
\bibliography{Tele-Parallel-Reference-file}

\end{document}